\documentstyle[prb,aps,eqsecnum]{revtex}
\begin{document}
\bibliographystyle{prsty}
\draft
\title{Josephson tunneling in high-$T_c$ superconductors}
\author{M. B. Walker and J. Luettmer-Strathmann}
\address{Department of Physics, University of Toronto,
Toronto, Ontario, M5S 1A7, Canada}
\date{received: December 19, 1995}
\maketitle
\begin{abstract}
This article describes the Josephson tunneling from 
time-reversal symmetry-breaking states and compares it with
that from time-reversal invariant  states for both
twinned and untwinned crystals and for both $c$-axis and 
basal-plane currents, in a model for orthorhombic YBCO. 
A macroscopic invariance group describing the superconducting
state of a twinned crystal is introduced and shown to provide 
a useful framework for the discussion of the results for twinned
crystals.
In addition, a ring geometry, which allows $s$-wave and $d_{x^2-y^2}$-wave
superconductivity in a tetragonal superconductor to be distinguished
on the basis of symmetry arguments only, is proposed and analyzed.
Finally, an appendix gives details
of the experimental Josephson tunneling evidence for a 
superconducting state of orthorhombic $ux^2+vy^2$ symmetry
in YBCO.
\end{abstract}


\section{Introduction}\label{intro}

Recently a number of Josephson tunneling 
experiments~\cite{wo93,br94,ma95b,wo95,va95b,ig94,ts94,ts95c,ts95d,%
ki95b,su94,ka95,dy94,wo95b,ch94} 
have been performed on the high-T$_c$  material
YBCO to investigate the symmetry of its order 
parameter.
Conventional superconductors all have order parameters with 
$s$-wave symmetry. On the other hand, many theoretical models
of high temperature superconductivity based on a description
of the pairing of the electrons in approximately tetragonal
copper-oxide layers (such as occur in YBCO) find superconducting
order parameters of symmetry $d_{x^2-y^2}$ (see 
Refs.~\onlinecite{da94} and \onlinecite{si95b} for reviews). 
Thus, experiments have tended to focus on distinguishing between
$s$ and $d_{x^2-y^2}$ symmetries for the superconducting order
parameter.

It is of interest to note that this focus on the YBCO order parameter
as a linear combination of $d_{x^2-y^2}$ and $s$-wave order parameters
can also be arrived at from a study of existing experimental results
only, without appeal to current microscopic theories. If the transition
to the superconducting state is continuous, then according to the 
Landau theory of second-order phase transitions, the superconducting 
order parameter must transform according to one of the eight irreducible
representations of the orthorhombic point group of YBCO. It is shown
in Appendix~\ref{appa}, that the observation of nonzero Josephson 
tunneling currents along all three principal crystallographic directions
in YBCO (e.g.\ as in Refs.~\onlinecite{wo93,br94,su94}) is sufficient
to establish that the YBCO superconducting order parameter is a 
linear combination of $d_{x^2-y^2}$- and $s$-wave components.

An important concept developed in this paper is that of the 
macroscopic symmetry (i.e.\ macroscopic invariance group) for the 
superconducting state of a twinned crystal. This macroscopic invariance
group (which is different from the microscopic invariance group 
describing the superconducting symmetry of an untwinned single crystal)
is the symmetry group determined by Josephson tunneling experiments
on twinned crystals. The majority~\cite{wo93,ma95b,wo95,va95b,ig94,ts94,%
ts95c,ts95d,ki95b} of Josephson
tunneling experiments on twinned YBCO crystals have determined
the macroscopic symmetry of the superconducting state to be the same
as that of a tetragonal $d_{x^2-y^2}$ superconductor.

According to Ref.~\onlinecite{wal95b} (see also Ref.~\onlinecite{an87}), 
two types of twin boundaries are possible within a Ginzburg-Landau 
model of YBCO in which the individual twins are in $d_{x^2-y^2}+s$ states:
the superconducting state can have either even or odd reflection symmetry
with respect to the twin boundary. Ref.~\onlinecite{wal95b} has shown
that the experimental result of macroscopic $d_{x^2-y^2}$ superconductivity
in twinned samples is equivalent to determining that the twin boundaries
in YBCO have odd reflection symmetry.

This leaves the question of the relationship between microscopic and 
macroscopic symmetry, which we discuss briefly below. As it turns out, 
a microscopic understanding of the energetics of the twin boundaries
is necessary to determine whether or not macroscopic $d_{x^2-y^2}$ symmetry
implies a microscopic symmetry which is predominantly $d$-wave
(see also Ref.~\onlinecite{va95b}). In this article, however, we  prefer
not to rely on microscopic arguments, and to consider all possibilities
consistent with the requirements of symmetry.

In principle, (e.g.\ assuming that difficulties concerning trapped
flux can be satisfactorily eliminated) a direct experimental determination
of whether $d_{x^2-y^2}$ or $s$ component of the order parameter is
dominant can be made by performing a SQUID experiment on an untwinned 
single crystal of YBCO in the geometry suggested by Sigrist and 
Rice~\cite{si95b}. Experiments on untwinned crystals have been performed
by Wollman {\it et al.}~\cite{wo93,wo95,va95b} and by Brawner and 
Ott~\cite{br94}, and indicate that the microscopic symmetry of the 
order parameter is predominantly $d$-wave.

The tricrystal ring experiments of Tsuei {\it et al.}~\cite{ts94,ts95c,ts95d%
,ki95b,ts96}
are of particular interest because of the precision with which problems
associated with trapped flux can be eliminated, and because one of these
experiments~\cite{ts96} has been carried out on 
Tl$_2$Ba$_2$CuO$_{6+\delta}$, which is tetragonal and therefore not 
open to questions of interpretation related to the existence of twins.
These experiments show convincingly rings containing a flux of one half
an elementary flux quantum in their ground state; such rings must 
contain either one or three $\pi$ Josephson junctions. Based on simplified
models for the angular dependence (on the angles between the Josephson
junction and the crystallographic axes) of the Josephson current, a plausible
argument is made that these results imply $d_{x^2-y^2}$ superconductivity.
However, it would not contradict the basic principles of physics if, 
for the experimental geometries used up to the present, $s$-wave 
superconductivity were to give rise to a ring whose ground state contained
half an elementary flux quantum. We therefore propose and analyze, in a 
section of our article below, a ring geometry where the observation
of one half a flux quantum in the ground state can exclude, on the 
basis of experimental measurements and symmetry arguments only, the 
possibility  that the superconducting state is $s$-wave.
The case of $d+is$ superconductivity (see below) is also discussed 
for this geometry.
 
The experiments of Ref.~\onlinecite{ch94} do not appear to yield
the $d_{x^2-y^2}$ macroscopic symmetry obtained by other Josephson
experiments performed on twinned crystals.
A possible explanation for this apparent inconsistency has been
given in Ref.~\onlinecite{mi94} in terms of the existence of 
vortices in the weak links. Further study of this experimental
configuration has indeed found magnetic flux trapped in the
grain boundary junctions~\cite{ki95}.

A remaining puzzle among the Josephson results on YBCO-Pb junctions
 concerns the 
$c$-axis critical currents. Because of the orthorhombicity of YBCO, 
there should be a nonzero $c$-axis Josephson current
from an untwinned single crystal 
 even if the
superconducting order parameter of YBCO is predominantly 
$d_{x^2-y^2}$-wave.
In the measurements of Ref.~\onlinecite{su94,ka95,dy94}, the $c$-axis
current was found to be nonzero both for twinned and untwinned crystals
of YBCO; the $I_{\rm c}R_{\rm n}$ product (where $I_{\rm c}$ is the 
critical current and $R_{\rm n}$ is the normal state resistance)
is approximately an order of magnitude smaller for films than it is
for single crystals~\cite{ka95}, with the highest value being
obtained for untwinned single crystals~\cite{dy94}.
On the other hand, efforts~\cite{wo95b} to detect Josephson tunneling
in cases where the Josephson junctions have a relatively high value
of $R_{\rm n}$ have not been successful, perhaps due to thermal
fluctuation effects. These varied results are not understood.
According to Ref.~\onlinecite{wal95b}, experimental measurements
of basal-plane Josephson currents indicate that the twin boundaries
have odd reflection symmetry; for such twin boundaries, the $c$-axis
Josephson currents from twinned crystals will average to zero, 
assuming the two types of twins are present with equal weight.
One possibility~\cite{wal95b} of accounting for a nonzero
$c$-axis current is that the twins are not present with equal weight
(perhaps as a result of strains); alternatives will be 
discussed below.

It is perhaps useful to introduce the idea of ``robustness'' of the
consequences of the macroscopic symmetry. The macroscopic invariance 
group of a twinned crystal is determined by assuming that all types
of twins are present with equal weight. However, it may happen that, 
due to strain for example, imbalances occur where one type of twin
has greater weight than another. A robust property is one which
is not changed by a small imbalance of this type.
It can be shown that the observation of  a $\pi$ phase shift in a basal-plane
corner SQUID experiment on YBCO is a robust property, whereas the 
expected zero value of the $c$-axis Josephson current from a twinned
sample having macroscopic $d_{x^2-y^2}$ symmetry is not. Thus, as 
noted in the previous paragraph, the $c$-axis results of 
Refs.~\onlinecite{su94,ka95,dy94} could be explained in terms of an 
unequal weighting of twin types.

As noted by Li {\it et al.}~\cite{li93} in a Ginzburg-Landau study
of mixed $s$ and $d_{x^2-y^2}$ superconductivity in an orthorhombic
superconductor such as YBCO, the phase transition to the superconducting
state will be, assuming that this transition is continuous, 
to a time-reversal invariant state called the $d+s$ state. They also
pointed out~\cite{li93} that a second transition to a different
superconducting state called the $d+e^{i\phi}s$ state is possible
at a lower temperature. This  $d+e^{i\phi}s$ state breaks time reversal
symmetry.
At present there is only one report~\cite{it94} of a second phase
transition in YBCO. It is however possible, particularly in twinned
samples, that such a second transition could be smeared out and 
not easily visible. 
The $d+e^{i\phi}s$ state is not stable in some microscopic
theories~\cite{xu95c}; a recent weak-coupling calculation~\cite{xu95c},
for example, finds that the $d+s$ is always stable relative to
the $d+e^{i\phi}s$ state.
Since there is at present no consensus concerning 
an appropriate  microscopic theory
of high temperature superconductivity, we do not exclude the $d+e^{i\phi}s$
state from our considerations
but explore its experimental consequences.

This article begins by discussing the time-reversal symmetry-breaking
$d+e^{i\phi}s$ state, the possible types of twin boundaries consistent
with this state, and the macroscopic symmetries of twinned crystals
having twins in the $d+e^{i\phi}s$ state. Some results relevant
to $d+s$ states, which are a special case ($\phi=0$) are included
for completeness.

Broken time-reversal symmetry in YBCO has been searched for in
optical experiments~\cite{sp90,la92}, $\mu$SR 
experiments~\cite{ki90}, and in Josephson tunneling experiments~\cite{ma95b}
and has not been found.
Nevertheless, because of the importance of the Josephson effect
as a method of determining  the order parameter symmetry, 
it is essential to have a clear understanding of the
experimental manifestations of broken time-reversal symmetry
for experiments performed on both untwinned and twinned crystals.
In related work, 
some effects of broken time-reversal symmetry at crystal grain
boundaries~\cite{si95} and at twin boundaries~\cite{si95c}
have recently been considered by other authors.

The Josephson current through a junction can be written as
\begin{equation}\label{i1}
I=I_c\sin(\phi-\phi_c) \, , 
\end{equation}
where $I_c>0$ is the Josephson critical current, and $\phi$ is the 
gauge invariant difference between the phases of the superconductors
on the two sides of the junction.
For normal junctions (e.g.\ junctions between conventional $s$-wave
superconductors) the characteristic phase of the junction  
can always be taken to be
$\phi_c=0$.
Experiments~\cite{wo93,br94,ma95b,wo95,va95b,ig94,ts94,%
ts95c,ts95d,ki95b,ts96} on high temperature
superconductors on the other hand, give evidence for 
circuits containing an odd number of 
$\pi$ junctions, 
i.e.\ junctions  
having a characteristic phase of $\phi_c=\pi$.
The calculations performed below show, for both twinned and untwinned
crystals, that a Josephson junction involving a superconductor in a
state which breaks time-reversal symmetry can have a characteristic 
phase $\phi_c$ which is neither zero nor $\pi$.
A similar conclusion concerning $\phi_c$ was previously reached in 
Ref.~\onlinecite{ts95c}.

In sections~\ref{trb} and \ref{tri}, the Josephson currents for twinned
samples are calculated by averaging over currents from the individual
twins; the result of this averaging is shown to be consistent with the 
expectations based on the appropriate macroscopic invariance group 
for the twinned sample.

In this article, it is assumed that the orthorhombic 
basal-plane unit cells
of neighboring twins are at right angles to each other, cf.\
Fig.~\ref{fig1}, whereas in fact 
they are rotated towards each other by a small angle.
This misorientation was considered carefully in Ref.~\onlinecite{wal95b}
and can be shown not to alter the results for Josephson tunneling
discussed here.

\section{Twin boundaries between states breaking 
time-reversal symmetry}\label{free}

The Landau free energy density describing coupled
$s$- and $d_{x^2-y^2}$-wave superconductivity in an orthorhombic
superconductor such as YBCO is~\cite{li93} 
\begin{eqnarray}
f_{\text{L}}&=&\alpha_s\vert\psi_s\vert^2 +\alpha_d\vert\psi_d\vert^2
\nonumber\\
&+&\frac{1}{2}\beta_s\vert\psi_s\vert^4 +\frac{1}{2}\beta_d\vert
\psi_d\vert^4+\beta_4\vert\psi_s\vert^2\vert\psi_d\vert^2
\nonumber\\
&+& (-1)^\epsilon \alpha(\psi_s\psi_d^\ast +\psi^\ast_s\psi_d)
+\beta_0(\psi_s^2\psi_d^{\ast 2}+\psi_s^{\ast 2}\psi_d^2)\nonumber\\
&+&(-1)^\epsilon
[\beta_1\vert\psi_d\vert^2+\beta_2\vert\psi_s\vert^2](\psi_s\psi_
d^\ast+\psi_s^\ast\psi_d) \, , \label{f1}  
\end{eqnarray}
where $\psi_s$ and $\psi_d$ denote the $s$- and $d_{x^2-y^2}$-wave
components of the order parameter, respectively, where 
$\alpha_s=\alpha_s'(T-T_{s0})$, $\alpha_d=\alpha_d'(T-T_{d0})$, 
$\alpha_s', \alpha_d'>0$, $T$ is the temperature, and  
where $\epsilon\in\{1,2\}$ indicates which of the two types of
neighboring twins is described, cf.\ Fig.~\ref{fig1}.
The appearance of the $(-1)^{\epsilon}$ factors in Eq.~(\ref{f1}), 
which is a generalization of the free energy presented in 
Ref.~\onlinecite{li93}, is related to
the assumption that the twin boundary between two twins 
is a plane of reflection symmetry of the underlying
crystal lattice~\cite{wal95b}.
With the definition of the coordinates  in Fig.~\ref{fig1},
the $d_{x^2-y^2}$ component of the order parameter, $\psi_{d}$, 
changes sign under reflection in the twinning plane.
To ensure that the free energy of the reflected state is the 
same as that of the original state factors of $(-1)^{\epsilon}$
are included in Eq.~(\ref{f1}).

For $\alpha=\beta_1=\beta_2=0$, with 
$\beta_s>0$, $\beta_d>0$, and $\beta_s\beta_d-(\beta_4-2|\beta_0|)^2>0$
for thermodynamic stability,  the free energy (\ref{f1}) describes
a tetragonal system undergoing a continuous phase transition at 
$T=\mbox{max}(T_{d0},T_{s0})$ to either a $d$-wave ($\psi_d\neq 0$, $\psi_s=0$)
or an $s$-wave ($\psi_s\neq 0$, $\psi_d=0$) superconducting state. 
A second transition to a mixed ($\psi_d\neq 0$, $\psi_s\neq 0$) state
occurs when $\alpha_s\beta_d=\alpha_d(\beta_4-2|\beta_0|)$ or
$\alpha_d\beta_s=\alpha_s(\beta_4-2|\beta_0|)$, respectively.
We define the relative phase $\phi_{sd}$ between $\psi_s$ and $\psi_d$ 
through
\begin{equation}\label{f1b}
\psi_d=|\psi_d|e^{i\phi_d} \, , \:\:\:
\psi_s=|\psi_s|e^{i(\phi_d+\phi_{sd})} \, .
\end{equation}
In the tetragonal case, $\phi_{sd}$ is
determined by the sign of $\beta_0$: 
$\phi_{sd}\in\{0,\pi\}$  for $\beta_0<0$, 
corresponding to a time-reversal invariant $d+s$ state, 
$\phi_{sd}\in\{\pi/2,3\pi/2\}$
for $\beta_0>0$, corresponding to a time-reversal symmetry-breaking $d+is$ 
state.

The terms with coefficients $\alpha$, $\beta_1$, and $\beta_2$ describe
the coupling of the $s$- and $d$-wave components of the order parameter
which exists in an orthorhombic crystal. 
A continuous phase transition from the normal to a superconducting
state occurs in the orthorhombic system when the smallest eigenvalue
of the quadratic form in Eq.~(\ref{f1}) goes through zero.
For $\alpha\ll 1$, i.e.\ weak orthorhombicity, 
and $\alpha_d<\alpha_s$ the transition is at temperature
$T_{d}\approx T_{d0}+\alpha^2/\alpha_s\alpha_d'$ to a state whose 
$d$-wave component $\psi_d$ grows with decreasing temperature 
approximately like
$|\psi_d|^2= \alpha_d'(T_d-T)/\tilde{\beta}_d$, where 
$\tilde{\beta}_d\approx\beta_d$. 
The $s$-wave component $\psi_s$ can be found as an expansion in powers
of $\psi_d$, and to lowest order is
\begin{equation}\label{f2}
\psi_s =(-1)^{\epsilon+1}(\alpha/\alpha_s)\psi_d .
\end{equation}
For this solution, $\psi_d$ and $\psi_s$ have the same phase
(modulo $\pi$) and the superconducting state is called a $d+s$ state.
For $|\alpha/\alpha_s|\ll 1$, $|\psi_s|\ll |\psi_d|$, 
and the superconducting state is said to be predominantly $d$-wave.
For $\alpha_s<\alpha_d$, on the other hand, a transition occurs at
$T_{s}\approx T_{s0}+\alpha^2/\alpha_d\alpha_s'$ to a predominantly
$s$-wave superconducting state. 

To determine the relative phase $\phi_{sd}$
between $\psi_s$ and $\psi_d$, cf.\ Eq.~(\ref{f1b}), 
we set $\alpha>0$ and, for simplicity, extend
the assumption of weak orthorhombicity by neglecting $\beta_1$ and 
$\beta_2$ in comparison with $\beta_0$. 
By minimizing the free energy~(\ref{f1})
we find, for $\beta_0>0$, the following conditions on the  
relative phase $\phi_{sd}$: 
\begin{equation}\label{f3}
\sin{\phi_{sd}}=0 ,
\end{equation}
or
\begin{equation} 
\cos{\phi_{sd}}=(-1)^{\epsilon +1}\frac{\displaystyle\alpha}
{\displaystyle 4\beta_0|\psi_s||\psi_d|} . \label{f4}
\end{equation}
Condition (\ref{f3}) yields a relative phase of $\phi_{sd}=0$ 
for $\epsilon =1$, $\phi_{sd}=\pi$ for $\epsilon =2$,
and a time-reversal invariant $d+s$ superconducting
state. This state is realized near the transition from the normal
state (see also Eq.~(\ref{f2})) and, as the temperature is lowered,
remains stable until $|\psi_d|$ and $|\psi_s|$ have grown sufficiently
to reduce the fraction on the right-hand side of Eq.~(\ref{f4}) to unity. 
At the temperature $T=T_{\phi}$ with 
$|\psi_s(T_{\phi})||\psi_d(T_{\phi})|=\alpha/4\beta_0$  
a second transition to a 
$d+e^{i\phi}s$ state takes place. For the $\epsilon=1$ twin,
for example, $|\phi_{sd}|$ increases continuously from zero for
temperatures below $T_{\phi}$. The values $\phi_{sd}=\pm |\phi_{sd}|$
correspond to two energetically equivalent states of twin $\epsilon=1$ 
that are related
to each other by the time-reversal operation ${\cal T}$. The $d+e^{i\phi}s$
state thus breaks time-reversal symmetry. 

In the case that $\beta_0<0$, on the other hand,
 the minimum of the free energy (\ref{f1}) is always
attained for $\phi_{sd}=0$ for $\epsilon=1$ and
$\phi_{sd}=\pi$ for $\epsilon=2$, so that
only one phase transition occurs and the superconducting state
is always the time-reversal invariant state  $d+s$. 

Our considerations so far have shown that the relative
phase $\phi_{sd}$ between the components of the order
parameter in a given twin is determined by the bulk free
energy density (\ref{f1}).
Contributions to the free energy
due to the interaction between 
the order parameters on opposite sides of a twin boundary
have to be considered to describe
the relative phases of the order parameter
components across a twin boundary.
For a predominantly $d$-wave superconductor the interface
free energy per unit area associated with interactions between twin 1 and
twin 2, cf.\ Fig.~\ref{fig1}, is to lowest order given by~\cite{wal95b}
\begin{equation}\label{f5}
f^{d}_{12}=B(\psi^*_{d1}\psi_{d2}+\psi_{d1}\psi^*_{d2}) ,
\end{equation}
where $\psi_{d1}$ and $\psi_{d2}$ denote the $d$-wave components
of the order parameter in twin 1 and 2, respectively.
For time-reversal invariant $d+s$ states, 
the free energy is minimized by $\psi_{d1}=\psi_{d2}$ 
($\psi_{d1}=-\psi_{d2}$) for 
$B<0$ ($B>0$), which implies that the overall superconducting states
have the form
\begin{equation}\label{f6a}
\Psi_u=(\psi_{d1},\psi_{s1},\psi_{d2},\psi_{s2})=
(\psi_{d},\psi_{s},\psi_{d},-\psi_{s})\, , \:\: B<0 \, , 
\end{equation}
and 
\begin{equation}\label{f6}
\Psi_g=(\psi_{d1},\psi_{s1},\psi_{d2},\psi_{s2})=
(\psi_{d},\psi_{s},-\psi_{d},\psi_{s})\, , \:\: B>0 \, .
\end{equation}
The states $\Psi_g$ and $\Psi_u$ satisfy $\sigma\Psi_g=\Psi_g$
and $\sigma\Psi_u=-\Psi_u$ and are thus even and odd with respect
to a reflection $\sigma$ in the twinning plane, cf.\ Fig.~\ref{fig5}.
An investigation of Josephson 
tunneling currents for time-reversal invariant
states has shown~\cite{wal95b} that the experimental results for YBCO are
consistent with odd symmetry under $\sigma$, so that the $d$-wave
component is the same on either side of the twin boundary.

The assumption that $\psi_{d1}=\psi_{d2}$ remains correct for the
time-reversal symmetry-breaking $d+e^{i\phi}s$ states
yields two possible states in twin 2 for a given state in twin 1:
\begin{eqnarray}\label{f7}
\Psi_1&=&e^{i\phi_d}(|\psi_d|,e^{i\phi_{sd}}|\psi_s|) , \\
\Psi_2&=&e^{i\phi_d}(|\psi_d|,-e^{i\phi_{sd}}|\psi_s|) , \label{f8} \\
\Psi_{\bar{2}}&=&e^{i\phi_d}(|\psi_d|,-e^{-i\phi_{sd}}|\psi_s|) . \label{f9}
\end{eqnarray}
Since $\Psi_2=-\sigma\Psi_1$, and since the state $\Psi_{\bar{2}}$
is related to the time reversed state of twin 1, defined as
$\Psi_{\bar{1}}=e^{i\phi_d}(|\psi_d|,e^{-i\phi_{sd}}|\psi_s|)$, by
$\Psi_{\bar{2}}=-\sigma\Psi_{\bar{1}}$, 
it is clear that all states have the same bulk free energy density.
Which of these two states of twin 2 prefers to be in contact with the 
given state of twin 1 is determined by the surface energy of 
interaction of twin 1 with twin 2.
Note that if the pair $(\Psi_1,\Psi_2)$ is favored, the overall state
is odd with respect to a reflection $\sigma$ in the twinning plane;
this type of twin boundary will be called an odd-symmetry twin boundary.
Otherwise, if the pair  $(\Psi_1,\Psi_{\bar{2}})$ is favored, the
overall superconducting state is invariant, to within a phase factor,
with respect to a reflection $\sigma$ followed by time reversal ${\cal{T}}$;
this second type of twin boundary will be called a ${\cal{T}}\sigma$ 
twin boundary. For a given material, only one type of twin boundary 
will be present; this will be the one which contributes the least
to the free energy.

\section{Macroscopic Symmetry}\label{macro}

Since a twinned sample will have some regions in which the twinning planes
are $(110)$ planes, and other regions where the twinning planes 
are $(\bar{1}10)$
planes, it is important to examine how the superconducting order is 
transferred from one region to another.
Consider the case where the twinning planes are odd-symmetry reflection
planes. Within a given twin, say twin 1, the superconductivity is
described by its order parameter $\Psi_1$, which is a mixture of 
$d_{x^2-y^2}$- and $s$-wave components (cf.\ Eq.~(\ref{f7})). 
This order parameter is invariant with respect to the orthorhombic
point group operations (rotations of $\pi$ about the orthorhombic 
principal axes and reflections in planes normal to the principal
axes) and hence, from the point of view of its point group symmetry, 
can be represented by a rectangle, as in Fig.~\ref{fig2}. 
(This same rectangle can also be used to indicate the relative 
orientation of the $a$ and $b$ axes in the different twins.)
At an odd-symmetry twin boundary, not only does the rectangle representing
the order parameter change its orientation on reflection across a 
twin boundary, but it changes sign. 
Thus, in Fig.~\ref{fig2}, the alternating plus and minus signs in the 
rectangles represent  the different signs of the order parameters
in the different twins.
Also, the interface between a region of $(110)$ twin boundaries
and a region of $(\bar{1}10)$ twin boundaries, as determined by 
experiment~\cite{zh93}, is shown in Fig.~\ref{fig2}. 
It is of interest that this interface is itself largely a twin 
boundary, with small breaks in this boundary. 
Thus, the superconducting order parameter should be transmitted
from a region of $(110)$ twin boundaries to a region of $(\bar{1}10)$
twin boundaries in exactly the same way as it is transmitted from one 
twin to another within a region of, say, $(110)$ twin boundaries.
In the case of ${\cal T}\sigma$ twin boundaries, the superconducting
order is passed from twin to twin and between regions of differently 
oriented twin boundaries in a similar way.

\subsection{Macroscopic invariance group for $d+e^{i\phi}s$ 
states}\label{free1}

Figure~\ref{fig2} also gives a simple picture of the macroscopic symmetry
of a highly twinned superconductor with odd-symmetry twin boundaries.
Note that Fig.~\ref{fig2}(b) is obtained from Fig.~\ref{fig2}(a)
by a rotation of $\frac{1}{2}\pi$ about the $c$-axis.
However, Fig.~\ref{fig2}(b) is also obtained from Fig.~\ref{fig2}(a)
by changing the signs of all of the order parameters, i.e.\ by 
multiplying the state of the entire system by minus one.
Thus, a rotation of this twinned crystal by $\frac{1}{2}\pi$ about the 
$c$-axis is equivalent to changing the sign of the overall order 
parameter. 
In other words, the superconducting state of the twinned crystal
is invariant under the operation $e^{i\pi}C_4$. 
Similarly, the superconducting state of the twinned crystal
can be shown to be invariant with respect to all of the operations of the 
direct product group $D_4^{(1)}(D_2)\times I$, where $I$ is the group
containing the identity and the inversion, and $D_4^{(1)}(D_2)$ is the
combined group $D_4^{(1)}(D_2)=(E,C_2,2U_2,2C_4e^{i\pi},2U'_2e^{i\pi})$ 
in the notation of Ref.~\onlinecite{vo85}.
The direct product group $D_4^{(1)}(D_2)\times I$
will be said to describe the macroscopic symmetry of superconductivity
in the twinned crystal. Note that this macroscopic symmetry group
lacks time reversal; the time-reversal operation creates a different 
superconducting state with the same free energy.

The macroscopic invariance group, $D_4^{(1)}(D_2)\times I$, of a twinned 
crystal with twins in $d+e^{i\phi}s$ states and with odd 
reflection-symmetry twin boundaries is thus the same, except for
the lack of the time-reversal operation, as the invariance group of the
$d_{x^2-y^2}$ superconducting state of an (untwinned) tetragonal single
crystal~\cite{vo85}. 
If the characteristics of the Josephson effect are determined by
the macroscopic invariance group of the twinned crystal, then we 
expect the $c$-axis Josephson current to be zero, and an offset flux
(cf.\ Eq.~(\ref{b02}) below) of $\frac{1}{2}\Phi_0$ in a corner SQUID
experiment. These characteristics of the Josephson effect for the type of
twinned-crystal superconductivity described in this paragraph are confirmed
by detailed twin averaging calculations in Sections~\ref{trb} and \ref{tri}.

The discussion of the case where the superconducting twin boundaries
are of type ${\cal T}\sigma$ can be carried out in a similar way.
In this case, the basic states of the two types of twins are $\Psi_1$
and $\Psi_{\bar{2}}$, as given by Eqs.~(\ref{f7}) and (\ref{f9}).
Detailed consideration of the effects of the elements of $D_4$ shows that
the macroscopic invariance group of this state of the twinned crystal
is $D_4^{(3)}(D_2)\times I$, where 
$D_4^{(3)}(D_2)=(E,C_2,2U_2,2{\cal T}e^{i\pi}C_4,2{\cal T}e^{i\pi}U'_2)$.
Thus, the macroscopic symmetry here is the same as that of the $d+is$ 
superconducting state of an (untwinned) tetragonal crystal.
Again, the time-reversal operation ${\cal T}$ by itself does not leave the
state invariant, but rather creates a different state with the same
free energy. 
The macroscopic symmetry here allows a nonzero $c$-axis Josephson current 
and an offset flux differing from $\frac{1}{2}\Phi_0$ in a corner SQUID
experiment; detailed calculations performed below indeed confirm these 
properties for a twinned crystal with ${\cal T}\sigma$ twin boundaries.

\subsection{Macroscopic invariance group for $d+s$ states}\label{free2}

A previous paper~\cite{wal95b} which presented detailed calculations
of the Josephson currents from an orthorhombically twinned crystal, with
twins in the time-reversal invariant $d+s$ state, to a conventional
superconductor, did not exploit the idea of the macroscopic symmetry
of the twinned crystal. Two types of twin boundaries are possible, 
giving rise to superconducting states that are either 
odd,  cf.\ Eq.~(\ref{f6a}), or even,  cf.\ Eq.~(\ref{f6}), with
respect to a reflection in the twinning plane.
The case of odd reflection-symmetry  twin boundaries
is the same as the case of odd reflection-symmetry twin 
boundaries studied above, except that $\phi_{sd}=0$ since the
twins are in time-reversal invariant states. Hence, the time-reversal
operation ${\cal T}$ is added to the invariance group, which becomes
$D_4^{(1)}(D_2)\times I\times R$, where $R$ contains the identity and
time-reversal; this is also the invariance group for $d_{x^2-y^2}$
superconductivity in a tetragonal superconductor.
For the case of even reflection-symmetry
twin boundaries, which can be represented schematically by assigning 
plus signs to all of the rectangles in Fig.~\ref{fig2},  
the macroscopic invariance group is $D_4\times I\times R$, 
which is also the invariance group for $s$-wave superconductivity
in a tetragonal superconductor.
Thus, in orthorhombically twinned crystals, with the twins in $d+s$ 
superconducting states, odd reflection symmetry twin boundaries give a 
macroscopic symmetry identical to that of a tetragonal $d_{x^2-y^2}$-wave 
superconductor, whereas even reflection-symmetry twin boundaries give a
macroscopic symmetry identical to that of a tetragonal $s$-wave 
superconductor. Clearly, as emphasized in Ref.~\onlinecite{wal95b}, the 
qualitative characteristics of the Josephson effect in twinned crystals, 
which are determined by the macroscopic symmetry of the superconducting
state, are a reflection of the symmetry of the twin boundaries, and have
nothing to do with whether the basic $d+s$ state of the twins is
predominantly  $d_{x^2-y^2}$-wave ($|\psi_d|\gg |\psi_s|$) or
predominantly $s$-wave ($|\psi_s|\gg |\psi_d|$).

\subsection{Relation between twin-boundary symmetry and microscopic
symmetry}\label{free3}

Given that Josephson experiments on twinned crystals of YBCO determine
directly the twin boundary symmetry, it is of interest to ask to what
extent this result can be used to infer the dominant microscopic 
order-parameter symmetry.

As noted above, if a phenomenological Ginzburg-Landau type of model 
is adopted for the twin boundaries (e.g.\ as in Eq.~(\ref{f5}) above 
or as in  Ref.~\onlinecite{an87}) they can have either even or odd 
reflection symmetry. Which case occurs depends on the signs and magnitudes
of certain parameters in the twin-boundary free energy. The fact
that the microscopic order parameter may be either predominantly
$d_{x^2-y^2}$-wave or predominantly $s$-wave places no restrictions
on these parameters at the phenomenological level.
Thus, within the framework of Ginzburg-Landau type models, the 
identification of the twin-boundary symmetry does not lead to information
about the dominant order parameter symmetry.

Now consider the odd-symmetry and even-symmetry twin boundaries as 
shown in Fig.~\ref{fig5}. For the odd-symmetry twin boundary, the 
$s$-wave component of the order-parameter changes sign, whereas the 
$d$-wave component is continuous. On the other hand, for an even-symmetry
twin boundary, the $d$-wave component of the order parameter changes
sign, while the $s$-wave component is continuous. If it is assumed
that either the $d_{x^2-y^2}$- or $s$-wave component of the order
parameter is completely dominant, and that the lowest free energy state
is achieved by having an order parameter that is continuous and 
varies as little as possible at the twin boundary, then predominantly
$d_{x^2-y^2}$-wave  ($s$-wave) superconductivity would lead to 
odd-symmetry (even symmetry) twin boundaries.  The argument that the 
dominant order parameter should be as continuous as possible seems 
reasonable, but there are at least hypothetical situations (see e.g.\
Ref.~\onlinecite{bu77}) for which it is not correct.

In YBCO, both infrared~\cite{ba95} and Josephson tunneling~\cite{su95}
measurements show a relatively strong anisotropy in the $ab$-plane
penetration depth. 
Furthermore, a recent microscopic analysis~\cite{at95}
 of the penetration depth
data indicates that the CuO chains play a significant role in the
superconductivity.
This suggests that neither the $d_{x^2-y^2}$ nor 
the $s$ component of the order parameter is completely dominant, but 
that both have significant amplitudes. 
For an odd-symmetry twin boundary, the assumption that the order 
parameters vary continuously across the twin boundary implies that
the $d_{x^2-y^2}$-wave component of the order parameter remains 
approximately uniform across the twin boundary. The $s$-wave
component, however, will have its bulk value $\psi_s$ at distances 
greater than $-\xi_s$ from the twin boundary
($\xi_s$ is the $s$-wave coherence length),  will go continuously
to zero and change sign at the twin boundary, and will tend to
the value $-\psi_s$ at distances greater than $+\psi_s$ from the
twin boundaries.

If the condensation 
energy per unit volume gained by having a nonzero $s$-wave order
parameter is $\epsilon_s$, then the free energy per unit area of
an odd reflection-symmetry twin boundary will, in this model, 
be approximately
\begin{equation}\label{free31}
f_{\text{odd}}=2\epsilon_s\xi_s+\delta f_{\text{odd}} .
\end{equation}
Here a quantity $\delta f_{\text{odd}}$ is assumed to include all
corrections necessary to arrive at the correct answer;
these corrections may be due to interactions between the $s$ and $d$-wave
components of the order parameter, effects of crystal structure and doping
level at the twin boundary giving condensation energies not equal to
the bulk value, the effect of disorder at the twin boundary which might
reduce the amplitude of the $d$-wave component of the order parameter, etc.
Similarly, in the case of an even reflection-symmetry twin boundary, 
the twin-boundary free energy per unit area for our simplified model
can be written 
\begin{equation}\label{free32}
f_{\text{even}}=2\epsilon_d\xi_d+\delta f_{\text{even}} .
\end{equation}

Within this basic model of the energetics of the twin boundary, 
a knowledge of the relative condensation energies, 
and of the relative coherence lengths of the $s$- and $d_{x^2-y^2}$-wave
components of the order parameter is required to determine whether the 
odd-symmetry or even-symmetry twin boundary is stable for a given
relative weight of $s$ and $d_{x^2-y^2}$ to the order parameter.
We are not aware of the existence of good quantitative estimates
of these quantities.

In summary, our view is that, while it may be possible to make 
appropriate assumptions which would suggest that the observation
of odd-symmetry twin boundaries implies a dominant $d_{x^2-y^2}$
component of the order parameter, the argument is not 
conclusive. There is thus an important role for experiments on
untwinned single crystals of YBCO~\cite{wo93,br94,wo95,va95b}, 
or on tetragonal 
crystals such as Tl$_{2}$Ba$_{2}$CuO$_{6+\delta}$~\cite{ts96}, 
since in these cases the complications due to twinning are 
eliminated.

Finally, it should be noted that related issues have also been
discussed in Ref.~\onlinecite{va95b}, where a somewhat different
point of view is taken.

\section{Josephson currents}\label{joseph}

This section establishes a relationship between the Josephson 
current across a junction and the phase derivative of the free
energy by extending the approach of Andreev~\cite{an87} 
(see Ref.~\onlinecite{si95b} for a review) to the case of a
multicomponent order parameter. In order to
calculate currents in the superconductor, 
gradient terms compatible with the orthorhombic symmetry
of the system are added 
to the free energy density (\ref{f1}). To lowest order, 
these are given by
\begin{equation}\label{j1}
F_{\text{grad}}=\int\! dV \, \frac{1}{2}\sum_{\alpha}
\sum_{\lambda,\mu}
\frac{\displaystyle 1}{\displaystyle m^{\alpha}_{\lambda,\mu}}
\left(p_{\alpha}\psi_{\lambda}\right)^*
\left(p_{\alpha}\psi_{\mu}\right) , 
\end{equation}
where $\alpha\in\{x,y,z\}$ and $\lambda,\mu\in \{d,s\}$.
The coefficients of the gradient terms are real, satisfy
$m^{\alpha}_{\lambda,\mu}=m^{\alpha}_{\mu,\lambda}$, and the
components $p_{\alpha}$ of the ``momentum'' operator are given by
\begin{equation}\label{j2}
p_{\alpha}=-i\hbar\nabla+
\frac{\displaystyle |e^*|}{\displaystyle c}A_{\alpha} \, ,
\end{equation}
where $A_{\alpha}$ is a component of the vector potential,
$c$ is the speed of light,
and $|e^*|$ is the absolute value of the charge of a Cooper pair.
The components $j_{\alpha}$ of the current are obtained from the 
variation of $F_{\text{grad}}$ with respect to $A_{\alpha}$, 
and are given by
\begin{equation}\label{j3}
j_{\alpha}=\sum_{\lambda,\mu} 
\frac{\displaystyle |e^*|}{\displaystyle 2 m^{\alpha}_{\lambda,\mu}}
\psi_{\lambda}^*p_{\alpha}\psi_{\mu} + \text{ c.c. }  .
\end{equation}
To derive a Josephson current-phase relation for our unconventional
superconductor,  
we generalize definition (\ref{f1b}) by writing 
$\psi_{\mu}=|\psi_{\mu}|e^{i(\phi_d+\phi_{\mu d})}$, where 
$\phi_{dd}=0$, and $\phi_{sd}$ is the relative phase defined in 
Eq.~(\ref{f1b}). The phase $\phi_{d}$ can vary with position but, 
as pointed out in the previous section, 
$\phi_{sd}$ is determined by the bulk free energy. 
Under the assumption that $|\psi_{\mu}|$ is independent of position, 
we find:
\begin{equation}\label{j3b}
j_{\alpha}=c_{\alpha}\left[ 
\frac{\displaystyle \partial \phi_d}{\displaystyle \partial x_{\alpha}} +
\frac{\displaystyle 2\pi}{\displaystyle \Phi_0}A_{\alpha}\right] \, , 
\:\: \Phi_0=\frac{\displaystyle hc}{\displaystyle |e^*|} \, ,
\end{equation}
where
\begin{equation}\label{j3c}
c_{\alpha}=\sum_{\lambda,\mu}
\frac{\displaystyle \hbar |e^*|}{\displaystyle m^{\alpha}_{\lambda,\mu}}
|\psi_{\lambda}||\psi_{\mu}| \cos{(\phi_{\mu d}-\phi_{\lambda d})}\, .
\end{equation}

The Josephson current across an interface 
between YBCO and a conventional superconductor like lead 
depends on the interface free energy associated with
interactions between the $s$- and $d$-wave
components $\psi_{\mu}, \mu\in\{s,d\}$ of the order parameter on
the YBCO side and the conventional ($s$-wave) order parameter 
$\psi_{\text{Pb}}$ on the lead side.
This interface free energy per unit area can be written as:
\begin{equation}\label{j5} 
f_I= \sum_{\mu} G_{\mu}({\bf n})
\left( \psi_{\mu}^*\psi_{\text{Pb}} + \text{ c.c. } \right) ,
\end{equation}
where $G_{\mu}({\bf n})$ is real and, in general, depends on the direction 
of the surface normal ${\bf n}$, 
relative to the YBCO crystallographic axes.
There are also contributions to the interface 
free energy that depend on $\psi_{\mu}$ and $\psi_{\text{Pb}}$
separately. Since these terms do not contribute to the tunneling
we neglect them here.

The minimization of the total interface free energy with respect
to variations in $\psi_{\lambda}^*$ yields
\begin{equation}\label{j6}
\sum_{\alpha,\mu}n_{\alpha}\frac{\displaystyle i\hbar}
{\displaystyle 2m_{\lambda,\mu}^{\alpha}}(p_{\alpha}\psi_{\mu})
=-G_{\lambda}({\bf n})\psi_{\text{Pb}}.
\end{equation}
Insertion of Eq.~(\ref{j6}) into (\ref{j3}) gives, for the Josephson
current ${\bf j}$ across an interface between YBCO and 
lead with normal vector ${\bf n}$,
\begin{equation}\label{j7}
{\bf n}\cdot{\bf j}=
\frac{\displaystyle -i|e^*|}{\displaystyle \hbar}
\sum_{\lambda}G_{\lambda}({\bf n})\psi_{\lambda}^*\psi_{\text{Pb}} 
+ \text{ c.c. } .
\end{equation}

As in the case of conventional superconductivity,
the current ${\bf j}$ is related to the derivative of the interface 
free energy $f_I$ with respect to the phase difference between the 
order parameters on both sides of the junction.
Substituting $\psi_{\text{Pb}}=|\psi_{\text{Pb}}|e^{i\phi_{\text{Pb}}}$,
and $\psi_d$ and $\psi_s$ from Eq.~(\ref{f1b}) into Eqs.~(\ref{j5})
and (\ref{j7}), 
and setting $\phi=\phi_d-\phi_{\text{Pb}}$ yields
\begin{equation}\label{j10}
\frac{\displaystyle \partial f_I(\phi)}{\displaystyle \partial \phi}=
\frac{\displaystyle \Phi_0 }{\displaystyle 2\pi c} {\bf n}\cdot{\bf j} \, .
\end{equation}

\section{\lowercase{$c$}-axis tunneling}\label{trb}

Consider as a first example 
Josephson tunneling along the $c$-axis between YBCO and lead.
The free energy per unit area $f_I$ describing the interaction between
the superconducting order parameters at the interface, which is normal
to the $c$-axis, can be written as:
\begin{equation}\label{c1}
f_I=-\left[ (-1)^{\epsilon +1}D\psi_d^*+S\psi_s^*\right]\psi_{Pb}
+\text{ c.c. } \, ,
\end{equation}
where the constants $(-1)^{\epsilon +1}D$ and $S$ are appropriate
realizations of the coefficients $G_{\mu}({\bf n})$ in
Eq.~(\ref{j5}). The coefficient $D$ is nonzero because of the 
orthorhombicity of YBCO.

We start with the interface energy $f_{I,1}$ for twin 1 for the
case that YBCO is in the time-reversal symmetry-breaking
$d+e^{i\phi}s$ state.
Let twin 1 be in the superconducting state $\Psi_1$ defined in
Eq.~(\ref{f7}) and let
$\psi_{\text{Pb}}=|\psi_{\text{Pb}}|e^{i\phi_{\text{Pb}}}$.
With the definitions
\begin{equation}\label{c2}
d=2D|\psi_d||\psi_{\text{Pb}}| \, , \:\:
s=2S|\psi_s||\psi_{\text{Pb}}| \, , \:\:
\phi=\phi_d-\phi_{\text{Pb}} \, ,
\end{equation}
we find 
\begin{equation}\label{c3}
f_{I,1}=-r_1\cos{(\phi-\phi_1)} \, ,
\end{equation}
with
$$
r_1=\sqrt{(d+s\cos{\phi_{sd}})^2+(s\sin{\phi_{sd}})^2}
$$
\begin{equation}\label{c4}
\sin{\phi_1}=-s\sin{\phi_{sd}}/r_1 \, , \:\:
\cos{\phi_1}=(d+s\cos{\phi_{sd}})/r_1 \, .
\end{equation}
Note that for the time-reversal symmetry-breaking state (i.e.\ 
$\phi_{sd}\neq 0$), the characteristic phase of the junction, $\phi_1$,
will be neither $0$ nor $\pi$.

If twin 2 is in the state $\Psi_2=-\sigma\Psi_1$, cf.\ Eq.~(\ref{f8}),
so that the twin boundary is of the odd symmetry type,
then the free energy $f_{I,2}$ follows from Eq.~(\ref{c3}) for
$f_{I,1}$ by replacing $\phi_{sd}\rightarrow \phi_{sd}+\pi$ and
$d\rightarrow -d$ with the result:
\begin{equation}\label{c5}
f_{I,2}=-r_1\cos{(\phi-\phi_1-\pi)}=-f_{I,1} \, .
\end{equation}
Hence, the average of the interface energy over the two types
of twins $(f_{I,1}+f_{I,2})/2$ vanishes which
implies, cf.\  Eq.~(\ref{j10}), that the Josephson currents also
average to zero. 
This result of the Josephson current averaging to zero for odd symmetry
twin boundaries is independent of the value of $\phi_{sd}$ and is thus
the same for the time-reversal invariant and the time-reversal 
symmetry-breaking states. Hence, 
the result obtained above from the ideas of the macroscopic
symmetry of a twinned crystal is confirmed:
the element $e^{i\pi}C_4$ of the macroscopic invariance group will not
allow $c$-axis Josephson currents.

Now suppose that twin 2 is in the state
$\Psi_{\bar{2}}=-\sigma\Psi_{\bar{1}}$, cf.\ Eq.~(\ref{f9}),
so that the twin boundaries are of the ${\cal{T}}\sigma$ type.
The free energy $f_{I,{\bar{2}}}$ is then obtained 
from Eq.~(\ref{c3}) with the
substitutions $\phi_{sd}\rightarrow -\phi_{sd}+\pi$ and
$d\rightarrow -d$, which yields
\begin{equation}\label{c6}
f_{I,{\bar{2}}}=-r_1\cos{(\phi+\phi_1-\pi)} \, .
\end{equation}
The average of the interface free energies over the twins is
then given by
\begin{equation}\label{c7}
\bar{f_I}=\frac{1}{2}(f_{I,1}+f_{I,{\bar{2}}})=-r_1\sin{\phi_1}\sin{\phi} \, .
\end{equation}
Together with Eq.~(\ref{j10}) this yields for the current, 
\begin{equation}\label{c8}
{\bf n}\cdot{\bf j}=-\frac{\displaystyle 2\pi c r_1}
{\displaystyle \Phi_0}\sin{\phi_1}\cos{\phi} \, .
\end{equation}
Hence, there is a partial cancellation of the Josephson currents
upon averaging over twins such that the critical
current $j_c$ is reduced from the value for the untwinned crystal
by a factor $|\sin{\phi_1}|$.
This result of partial but not total cancellation of the Josephson
currents in twinned crystals resembles to some extent what has 
been observed in experiments, namely, that the highest
values of the $I_{\rm c}R_{\rm n}$ product have been observed
in untwinned crystals~\cite{dy94}, and that smaller values
(which are however nonzero) are observed in twinned 
crystals~\cite{su94,ka95,dy94} 
(see however Ref.~\onlinecite{wo95b}).
Note finally, that the macroscopic invariance group found above for 
the case of ${\cal T}\sigma$ twin boundaries allows a nonzero
Josephson current, in agreement with what has been found here.

\section{Basal plane Josephson Currents}\label{tri}

The experimental study of the basal plane Josephson currents is often
carried out using the corner SQUID 
geometry~\cite{wo93,br94,ma95b,si95b} of Fig.~\ref{fig3}.
The total current $I_{\text{total}}$, which is the sum of the currents
through junctions $A$ and $B$ is
\begin{equation}\label{b01}
I_{\text{total}}=I_A\sin(\phi_A-\phi_{cA})+I_B\sin(\phi_B-\phi_{cB}) \, .
\end{equation}
The difference in the gauge invariant phase differences for the two
junctions is $\phi_A-\phi_B=2\pi\Phi/\Phi_0$, where $\Phi$ is the flux
through the SQUID and $\Phi_0$ is the elementary flux quantum.
The average phase $(\phi_A+\phi_B)/2$ is a free variable which adjusts
itself so that the total current $I_{\text{total}}$ equals the current
fed into the SQUID. The maximum value of $I_{\text{total}}$
with respect to variations of $(\phi_A+\phi_B)/2$ is
\begin{eqnarray}\label{b02}
\lefteqn{I_{\text{max}}=(I_A+I_B)} \nonumber \\
&&\times\left\{
\epsilon^2+(1-\epsilon^2)\cos^2\left[ \frac{\pi}{\Phi_0}
(\Phi-\Phi_{\text{offset}}) \right] \right\}^{1/2} \, ,
\end{eqnarray}
where
\begin{equation}\label{b03}
\Phi_{\text{offset}}=\frac{\phi_{cA}-\phi_{cB}}{2\pi}\Phi_0 \, ,
\end{equation}
and
\begin{equation}\label{b04}
\epsilon=(I_A-I_B)/(I_A+I_B) \, .
\end{equation}
Note that $I_{\text{max}}$ is periodic in the flux $\Phi$ with 
period $\Phi_0$. The quantity $\Phi_{\text{offset}}$ is the flux
at which $I_{\text{max}}$ has its largest value.
For two normal junctions $\Phi_{\text{offset}}=0$, whereas 
for a SQUID with one normal and one $\pi$ junction, $\Phi_{\text{offset}}$
has the value $\Phi_0/2$~\cite{wo93,wo95,si95b}.
It is now shown that $\Phi_{\text{offset}}$ can have other values
when the superconducting state is not time-reversal invariant.

Consider a plane Josephson junction between twin 1 of YBCO and 
an $s$-wave superconductor such as lead. The junction is normal to the 
YBCO basal plane and its free energy $f_{I,1}$ per unit area is
\begin{equation}\label{b1}
f_{I,1}(\theta)=
-\left[ D(\theta)\psi_{d1}^*+S(\theta)\psi_{s1}^*\right]\psi_{Pb}
+\text{ c.c. } \, ,
\end{equation}
where $\theta$ is the angle between the normal to the interface ${\bf n}$
and the basis vector ${\bf a_1}$.
By reflecting the state $(\psi_{d1},\psi_{s1})$ of twin 1 with interface 
angle $\theta$ in the twin boundary,
a state $(\psi_{d2},\psi_{s2})=(-\psi_{d1},\psi_{s1})$   
for twin 2 is obtained which has interface angle $-\theta +\pi/2$
and the same free energy.  Hence the interface free energy for twin 2
is given by
\begin{eqnarray}\label{b2}
f_{I,(2)}(\theta)&=&
-\left[ -D(-\theta+\frac{\pi}{2})\psi_{d2}^*+
S(-\theta+\frac{\pi}{2})\psi_{s2}^*\right]\psi_{Pb}  \nonumber \\
& & +\text{ c.c. } \, . 
\end{eqnarray}
Orthorhombicity implies that
\begin{eqnarray}\label{b3}
&D(\theta+\pi)=D(-\theta)=D(\theta)& \, , \nonumber \\
&S(\theta+\pi)=S(-\theta)=S(\theta)& \, . 
\end{eqnarray}
With the aid of definition~(\ref{f7}) for the superconducting state,
the interface free energy for twin 1 can be written as
\begin{equation}\label{b4}
f_{I,1}(\theta)=-r(\theta)\cos(\phi-\phi_c(\theta)) \, ,
\end{equation}
where $\phi=\phi_d-\phi_{\text{Pb}}$, and where
\begin{eqnarray}\label{b4a}
& & r(\theta)=\sqrt{\tilde{d}^2(\theta)+\tilde{s}^2(\theta)} \, ,
\nonumber \\
& & \cos{\phi_c(\theta)}=\frac{\tilde{d}(\theta)}{r(\theta)} \, , \:\:
\sin{\phi_c(\theta)}=-\frac{\tilde{s}(\theta)}{r(\theta)} \, , \:\:
\end{eqnarray}
with
\begin{eqnarray}\label{b7}
\tilde{d}(\theta)&=&
2(D(\theta)|\psi_d|+S(\theta)|\psi_s|\cos{\phi_{sd}})|\psi_{\text{Pb}}|
\, , \nonumber \\
\tilde{s}(\theta)&=&
2S(\theta)|\psi_s||\psi_{\text{Pb}}|\sin{\phi_{sd}} \, .
\end{eqnarray}

Combining Eqs.~(\ref{j10}) and (\ref{b4}) gives, for an untwinned 
single crystal, a Josephson current
in the standard form of Eq.~(\ref{i1}), with the characteristic phase
of the junction $\phi_c(\theta)$ given by Eq.~(\ref{b4a}).
For the time-reversal invariant $d+s$ state, $\phi_{sd}=0$ and the 
characteristic phase of the junction will be either zero or $\pi$.
Results appropriate to a tetragonal superconductor in a 
$d_{x^2-y^2}$-wave state are obtained from the above formula by 
taking $S(\theta)=0$ and $D(\theta+\frac{1}{2}\pi)=-D(\theta)$.
This gives $\phi_c(\theta+\frac{1}{2}\pi)=\phi_c(\theta)+\pi$ and
an offset flux in a corner SQUID experiment of 
$\Phi_{\text{offset}}=\frac{1}{2}\Phi_0$.
For an orthorhombic crystal such as YBCO, there is however no 
symmetry-required relationship between $D(\theta+\frac{1}{2}\pi)$
and  $D(\theta)$, and also $S(\theta)$ is nonzero in general.
Nevertheless, provided the quantities $\tilde{d}(\theta+\frac{1}{2}\pi)$
and $\tilde{d}(\theta)$ have appropriate signs, the $d+s$ state
($\phi_{sd}=0$) will yield the value 
$\Phi_{\text{offset}}=\frac{1}{2}\Phi_0$ in a corner SQUID experiment.
Under these conditions, the superconducting state of orthorhombic
YBCO may be said to exhibit predominantly $d_{x^2-y^2}$-wave behavior.
For the time-reversal symmetry-breaking $d+e^{i\phi}s$ state,
$\phi_{sd}\neq 0$, and $\phi_c(\theta)$ and $\phi_c(\theta+\frac{1}{2}\pi)$
differ from zero and $\pi$, and $\Phi_{\text{offset}}$ differs from
zero or $\frac{1}{2}\Phi_0$.
If $|\psi_s|\ll |\psi_d|$ and/or $\phi_{sd}\ll\frac{1}{2}\pi$, 
this difference may be small and difficult to detect experimentally.
 
To describe basal-plane tunneling between a twinned sample of 
YBCO and lead, the interface free energy $f_I$ is averaged over the two
types of twins. The interface free energy $f_{I,\bar{2}}(\theta)$ 
of twin 2, cf.\ Eq.~(\ref{b2}), 
with the superconducting state $\Psi_{\bar{2}}$, given by Eq.~(\ref{f9}),
can be expressed in the form of Eq.~(\ref{b4}):
\begin{equation}\label{b14}
f_{I,\bar{2}}=-r(\theta+\pi/2)\cos(\pi+\phi+\phi_c(\theta+\pi/2)) \, ,
\end{equation}
which yields for the average $\bar{f}_I(\theta)=(f_{I,1}+f_{I,{\bar{2}}})/2$
\begin{equation}\label{b15}
\bar{f}_{I}(\theta)=
-r_{\bar{2}}(\theta)\cos(\phi-\phi_{\bar{2}}(\theta)) \, ,
\end{equation}
where
\begin{eqnarray}\label{b16}
& & r_{\bar{2}}(\theta)=\sqrt{\tilde{d}_d^2(\theta)+\tilde{s}_s^2(\theta)} \, ,
\nonumber \\
& &
\cos{\phi_{\bar{2}}(\theta)}=\frac{\tilde{d}_d(\theta)}{r_{\bar{2}}(\theta)} 
\, , \:\: 
\sin{\phi_{\bar{2}}(\theta)}=-\frac{\tilde{s}_s(\theta)}{r_{\bar{2}}(\theta)}
\, . \:\:
\end{eqnarray}
Here the $\theta$-dependent functions have been separated into
even  ($s$-wave like) and odd ($d$-wave like) 
terms under  $\theta\rightarrow\theta +\pi/2$ by 
\begin{eqnarray}\label{b8}
x(\theta)&=&\frac{1}{2}\left[ x(\theta)+x(\theta+\pi/2)\right] +
\frac{1}{2}\left[ x(\theta)-x(\theta+\pi/2)\right] \nonumber \\
&\equiv& 
x_s(\theta)+x_d(\theta) \, , 
\end{eqnarray}
where $x(\theta)$ stands for $\tilde{d}(\theta)$ or $\tilde{s}(\theta)$. 
Also, since the two twins have been taken to be in states $\Psi_1$ and 
$\Psi_{\bar{2}}$, the twin boundaries are what have been called above
${\cal{T}}\sigma$ twin boundaries. The above results 
(in particular Eq.~(\ref{b16})) show that for a twinned sample with 
${\cal{T}}\sigma$ twin boundaries the characteristic phases of the Josephson
junctions satisfy 
$\phi_{\bar{2}}(\theta+\frac{1}{2}\pi)=-\phi_{\bar{2}}(\theta)+\pi$.
This yields an offset flux in a corner SQUID experiment of 
\begin{equation}\label{b17}
\Phi_{\text{offset}}=\frac{\phi_{\bar{2}}(\theta)}{\pi}\Phi_0+
\frac{1}{2}\Phi_0 \, .
\end{equation}
The macroscopic invariance group of a crystal with ${\cal T}\sigma$ twin
boundaries, which is the same as that of a tetragonal crystal in a 
$d+is$ state, does not constrain the offset flux to be $0$ or 
$\frac{1}{2}\pi$, in agreement with Eq.~(\ref{b17}).
A value of $\Phi_{\text{offset}}={(0.98\pm 0.05)}\Phi_0/2$ 
is obtained in experiments~\cite{ma95b} on YBCO. 
Thus, either the YBCO twins are not in states of broken time-reversal
symmetry (i.e.\ $\phi_{sd}=0$), or the time-reversal symmetry-breaking
is small (more explicitly, $\tilde{s}_s\ll \tilde{d}_d$), 
or the twin boundaries are not of the type ${\cal T}\sigma$, or
time-reversal domains (see below) play a role.

Alternatively, if the twin boundaries are of the odd reflection symmetry
type, then twin 2 is in state $\Psi_2$ given by Eq.~(\ref{f8}), and 
Eq.~(\ref{b16}) holds but with $\tilde{s}_s(\theta)$ replaced by
$\tilde{s}_d(\theta)$.
This yields $\phi_2(\theta+\frac{1}{2}\pi)=\phi_2(\theta)+\pi$ for the
critical phase of the Josephson junction, and 
$\Phi_{\text{offset}}=\frac{1}{2}\Phi_0$.
Thus, for odd-symmetry twin boundaries, the same value of the offset
flux is obtained for the time-reversal invariant $d+s$ state and for the 
time-reversal symmetry-breaking $d+e^{i\phi}s$ state. 
It is therefore not possible to determine whether or not YBCO is in 
a time-reversal symmetry-breaking state from the experimental results
of Ref.~\onlinecite{ma95b} alone.
These results can again be qualitatively understood in terms of the
macroscopic invariance group of the twinned crystal, which for this
case is the same, except for the absence of time reversal, as that of
the $d_{x^2-y^2}$ superconducting state of a twinned crystal.

There remains the question of the existence of domains which are time-reversal
images of one another. Given a time-reversal symmetry-breaking
state of a twinned crystal, a different state of
the crystal having the same free energy can be obtained by applying the
time-reversal operation. The various equations describing this state
can be obtained from those presented in this and the preceding Section by
reversing the sign of $\phi_{sd}$. It may be that these two different 
states exist in different regions of the same crystal, in which case the 
crystal is said to contain time-reversal domains. If these time-reversal
domains are sufficiently small, so that it is necessary to average over them
to obtain the Josephson currents through a junction, then the behavior
characteristic of broken time-reversal symmetry will disappear. This makes
sense since a crystal containing a random distribution of domains which
are time-reversed images of each other is macroscopically invariant under
time reversal. Thus, the $c$-axis Josephson current found in
the case of ${\cal{T}}\sigma$ twin boundaries, cf.\ Eq.~(\ref{c8}), 
vanishes when averaged over time-reversal domains (since $\phi_1$
changes sign in going from a given domain to its time reversed image).
Similarly, an average of the basal plane Josephson currents over 
time-reversal domains will yield junctions whose characteristic 
phases can only be either 0 or $\pi$.

\section{Proposed geometry for ring experiment}\label{ring}

Recently, a number of order parameter symmetry determinations 
have been carried out using tricrystal ring 
magnetometry~\cite{ts94,ts95c,ts95d,ki95b,ts96}. 
These experiments have measured
the flux through a superconducting ring containing three Josephson 
junctions. 
The observation of a spontaneous magnetization of the ring corresponding
to a magnetic flux through the ring of $\frac{1}{2}\Phi_0$, where 
$\Phi_0$ is the elementary flux quantum, shows that one or perhaps
three of the Josephson junctions have a characteristic phase of $\pi$.
These experiments are impressive in eliminating the possible influence
of spurious trapped flux, and in the precision with which the spontaneous
flux is measured.

The identification of a Josephson junction having a characteristic 
phase of $\pi$ was taken to be evidence for $d_{x^2-y^2}$ symmetry.
This conclusion was based on a simplified model~\cite{si92} of the 
angular dependence (on the angles between the Josephson junction
and the crystallographic axes) of the Josephson current expected 
for $d_{x^2-y^2}$ superconductivity. The use of this simplified model
is somewhat analogous to the description of an electronic band structure
in  a tight binding model restricted to nearest neighbor interactions only.
Such a model is a good first guess, but its reliability is in fact
unknown. A complete parametrization of the Josephson current is given
in Appendix~\ref{appb}, where it is seen that, even to ``lowest order'', 
the parametrization of Ref.~\onlinecite{si92} does not retain all terms.
It should also be stated that, in its study of YBCO, 
Ref.~\onlinecite{ts95c}  examined three different geometries all of which
provided results consistent with the simplified $d$-wave model, 
and inconsistent with other possibilities which they raised.
For the geometries studied in 
Refs.~\onlinecite{ts94,ts95c,ts95d,ki95b,ts96}, however,
 it would be extremely
difficult to reach definite conclusions concerning the symmetry if a 
general model (similar to that developed in Appendix~\ref{appb})
for the Josephson current were adopted.
For this reason, we propose below a different ring geometry which 
can distinguish between $d_{x^2-y^2}$ and $s$ wave superconductivity
on the basis of symmetry arguments only, and without reference to 
simplified models of the Josephson junction.

It should also be noted that, as was the case with the corner SQUID
experiments discussed above, the symmetry which is determined in
these ring experiments on twinned crystals of YBCO is the
{\it macroscopic} symmetry. Thus, a determination that the macroscopic
symmetry is $d_{x^2-y^2}$ would lead directly to the conclusion that
the twin boundaries have odd reflection symmetry, while further 
arguments are needed to determine the {\it microscopic} symmetry of the
order parameter, cf. Section~\ref{free3}. 
Experiments on tetragonal
crystals, such as those of Ref.~\onlinecite{ts96} on 
Tl$_2$Ba$_2$CuO$_{6+\delta}$ are thus of special importance in that the 
complicating effect of twinning is eliminated.

In view of the above, it would be useful to have an experimental test
for $s$-wave versus $d_{x^2-y^2}$-wave 
superconductivity in tetragonal crystals
which is dependent on symmetry arguments only. The experimental geometry of
Fig.~\ref{fig4} provides such a test.
The superconductors labeled $S1$ and $S2$ are the same 
(e.g.\ {Tl$_2$Ba$_2$CuO$_{6+\delta}$}), but $S2$ has its $a$ and $b$ axes
oriented along the $x$ and $y$ axes of the figure, whereas the $a$ and $b$
axes of $S1$ are rotated from $x$ and $y$ through an angle of $\pi/4$ about
the $c$ axis.

Considerations of gauge invariance and time-reversal symmetry show that
the sum of the free energies of the Josephson junctions $A$ and $B$ 
(to terms quadratic in the order parameter) has the form
\begin{equation}\label{r1}
F=c_A(\Psi_{1A}\Psi_{2A}^*+\Psi_{1A}^*\Psi_{2A}) 
+c_B(\Psi_{1B}\Psi_{2B}^*+\Psi_{1B}^*\Psi_{2B}) \, ,
\end{equation}
where $\Psi_{1A}$ is the order parameter of superconductor $S1$ at Josephson
junction $A$, etc.\ , and where $c_A$ and $c_B$ are real.
If the superconductivity is $s$-wave, the invariance of the free energy
with respect to rotations of the ring by $\pi$ about the $x$ axis yields
the result $c_A=c_B$; in this case both junctions are either $\pi$ junctions
or normal junctions, and the ring will not exhibit a spontaneous magnetic
moment in its ground state.
On the other hand, if the superconductivity is $d$-wave, the same symmetry
argument requires that $c_A=-c_B$; hence in this case one junction is
normal while the other is a $\pi$ junction and the ring will exhibit a 
spontaneous moment in its ground state. The flux associated with this 
moment is $\frac{1}{2}\Phi_0$ in the limit $(LI_c/c)\gg\Phi_0/(2\pi)$,
where $I_c$ is the critical current and $L$ is the self-inductance of 
the ring~\cite{si95b,ts96,bu77}.

Note that in the geometry of Fig.~\ref{fig4}
both $\alpha=0$ and $\alpha=\pi/4$ yield $c_A=c_B=0$ for
$d_{x^2-y^2}$-wave superconductivity so that these angles would not
be appropriate choices for the experiment. 
Hence, values of $\alpha$ such as $\alpha\sim\pi/8$ should
allow a clear distinction to be made,
on symmetry grounds only, between $s$-wave and $d_{x^2-y^2}$-wave
superconductivity.

In addition, it has been pointed out to us~\cite{ts96p} that the 
model of Ref.~\onlinecite{ts95c} for the strong disordered limit
of the Josephson junction yields $c_A=c_B=0$ for all $\alpha$;
thus, it would be prudent to have the least possible disorder 
in the junction.

We now give a detailed analysis  which includes 
 three possible types of superconductivity 
for a tetragonal system, cf.\ 
Section~\ref{free}, namely $d_{x^2-y^2}$, $s$ and $d+is$.
Assume that in general the superconductivity can be described in 
terms of two complex order parameters, $\psi_d$ and $\psi_s$, 
of $d_{x^2-y^2}$ and $s$ symmetry, respectively. Then the 
free energy of junction $A$ has the form
\begin{equation}\label{r2}
F_A=G_d\psi_{d1A}\psi_{d2A}^* +G_s\psi_{s1A}\psi_{s2A}^*  
+G_{sd}\psi_{s1A}\psi_{d2A}^*  + G_{ds}\psi_{d1A}\psi_{s2A}^* 
+ \text{ c.c.} \, ,
\end{equation}
where $\psi_{d1A}$ is the order parameter of $d_{x^2-y^2}$ symmetry
in superconductor $S1$ at junction $A$, etc.\ , and where all coefficients
$G$ are real. The free energy for junction $B$ is the same, except
that $G_d\rightarrow -G_d$, $G_{ds}\rightarrow -G_{ds}$, and the 
subscript $A$ is replaced by subscript $B$ everywhere. 
For the $d+is$ state, the order parameter at junction $A$ takes the form
\begin{eqnarray}\label{r3}
\psi_{d1A}=|\psi_{d1A}|e^{i\phi_{d1A}} \, , & &
\psi_{d2A}=|\psi_{d2A}|e^{i\phi_{d2A}} \, , \nonumber \\
\psi_{s1A}=|\psi_{s1A}|e^{i(\phi_{d1A}+\frac{1}{2}\pi)} \, , & &
\psi_{s2A}=\lambda|\psi_{s2A}|e^{i(\phi_{d2A}+\frac{1}{2}\pi)} \, , 
\end{eqnarray}
where $\lambda=\pm 1$ accounts for the freedom to choose either of the two
$d\pm is$ states in superconductor $S2$. The order
parameter at junction $B$ is similarly expressed. 
Further analysis shows that the sum of the free energies of junctions $A$
and $B$ can be written as
\begin{equation}\label{r4}
F=-\frac{\Phi_0}{2\pi c}I_A\cos(\phi_A-\phi_{cA})
-\frac{\Phi_0}{2\pi c}I_B\cos(\phi_B-\phi_{cB}) \, ,
\end{equation}
where $I_A$, $I_B$ are positive.
For pure $d_{x^2-y^2}$-wave superconductivity, $\phi_{cA}-\phi_{cB}=\pm\pi$.
For pure $s$-wave superconductivity, $\phi_{cA}-\phi_{cB}=0$.
For $d+is$ superconductivity, $\phi_{cA}$ and  $\phi_{cB}$ are
(as a result of nonzero $G_{sd}$ and $G_{ds}$) different from zero 
(mod $\pi$). The quantities $\phi_A=\phi_{d1A}-\phi_{d2A}$ and 
$\phi_B=\phi_{d1B}-\phi_{d2B}$, when generalized to the usual gauge
invariant form, yield
\begin{equation}\label{r5}
\frac{1}{2}(\phi_A-\phi_B)=\frac{\pi\Phi}{\Phi_0} \, ,
\end{equation}
where $\Phi$ is the flux through the ring. The quantity 
$\frac{1}{2}(\phi_A+\phi_B)$ is at present undetermined; minimizing $F$
with respect to it yields
\begin{eqnarray}\label{r6}
F&=&-\frac{\Phi_0}{2\pi c}(I_A+I_B)\left\{ \epsilon^2+(1-\epsilon^2)
\cos^2\left[\frac{\pi}{\Phi_0}(\Phi-\Phi_{\text{offset}})\right]\right\}^{1/2}
\nonumber \\
& & \mbox{}+\frac{1}{2c^2}LI^2 \, , 
\end{eqnarray}
where
\begin{equation}\label{r7}
\Phi_{\text{offset}}=\frac{\phi_{cA}-\phi_{cB}}{2\pi} \Phi_0 \, , 
\end{equation}
and
\begin{equation}\label{r8}
\epsilon=\frac{|I_A-I_B|}{I_A+I_B} \, .
\end{equation}
Here, a quantity $LI^2/(2c^2)$, representing the magnetic field energy
associated with the current $I$ circulating in the ring, has been added
to $F$. The flux through the ring due to this current is
$\Phi_I=LI/c$, where $L$ is the ring self-inductance, and the total flux
through the ring is $\Phi=\Phi_I+\Phi_{\text{ex}}$, where $\Phi_{\text{ex}}$
is the flux due to external sources.

The ground state free energy of the ring in zero external flux, 
$\Phi_{\text{ex}}=0$, is found by minimizing $F$ with respect to $\Phi$.
For sufficiently large values of the parameter $L(I_A+I_B)2\pi/(c\Phi_0)$, 
the term in $LI^2$ can be treated in perturbation theory. 
Also, without loss of generality, we can assume 
$0\leq(\phi_{cA}-\phi_{cB})\leq\pi$. From 
these assumptions it follows that the flux through the ring in its 
ground state is $\Phi_{\text{offset}}$. Thus, the flux through the ring
in its ground state is zero for an $s$-wave superconductor, 
$\frac{1}{2}\Phi_0$ for a $d_{x^2-y^2}$-wave superconductor, 
and $x\Phi_0$, where $0<x<\frac{1}{2}$,
for a $d+is$-wave superconductor. 
It should be noted that the minimum flux of $x\Phi_0$ in the ground
state for the $d+is$ state occurs for the flux in a particular direction;
the minimum flux in the other direction is $(1-x)\Phi_0$, this asymmetry
being a reflection of the broken time-reversal symmetry.
(It has been noted previously in Ref.~\onlinecite{ts95c} that 
the $d+is$ state would lead to a ground state flux in a ring of $x\Phi_0$
where $x$ is neither $0$ nor $\frac{1}{2}$.)
If the parameter $L(I_A+I_B)2\pi/(\Phi_0 c)$ is not large, the ground
state flux through the ring will be less than $\frac{1}{2}\Phi_0$
for a $d_{x^2-y^2}$ superconductor, 
as in the case of a single-junction ring~\cite{ts96,bu77}, 
so that care will have to be taken to distinguish experimentally
between this situation, and the one where the departure of the ground
state flux from $\frac{1}{2}\Phi_0$ is due to broken time-reversal
symmetry.
These results, for the geometry of Fig.~\ref{fig4}, depend on symmetry
arguments only, and do not rely for their validity on elementary models
of the angular dependence of the Josephson critical currents.

Finally, note that although the observation of a ring containing 
flux $\frac{1}{2}\Phi_0$ in the proposed geometry would definitively
rule out $s$-wave symmetry for the superconductivity, and would be 
consistent with $d_{x^-y^2}$-wave superconductivity, it would not
definitively establish the superconductivity as being $d_{x^2-y^2}$-wave;
$d_{xy}$ superconductivity would also be consistent with such a result.
Other arguments, analogous to those of the Appendix, could however
definitively establish a $d_{x^2-y^2}$ state.

\section{Conclusions}\label{conc}

Since YBCO is orthorhombic, the order parameter describing its 
superconductivity is expected to transform according to one of
the irreducible representations of the orthorhombic point group.
The experimental 
observation~\cite{wo93,br94,ma95b,wo95,va95b,ig94,%
ts94,ts95c,ts95d,ki95b,su94,ka95,dy94,wo95b}
of nonzero bilinear Josephson critical currents along at least
two of the three principal orthorhombic axes in tunneling from
YBCO to a conventional $s$-wave superconductor is sufficient
to identify the symmetry of the order parameter as being of the
$ux^2+vy^2$ ($A_{1g}$) type (see Appendix). This provides the experimental
justification for considering YBCO superconductivity as coming
from a mixture of $d_{x^2-y^2}$-wave and $s$-wave contributions, 
which are the two tetragonal symmetry types compatible with 
orthorhombic $A_{1g}$ symmetry.

In a model~\cite{li93} 
of mixed $s$ and $d_{x^2-y^2}$ superconductivity in YBCO, 
there is a transition to a time-reversal invariant $d+s$ state at the 
normal-to-superconducting transition temperature $T_c$, followed by
a possible transition to a time-reversal symmetry-breaking 
$d+e^{i\phi}s$ state at a lower temperature.
Corresponding to a $d+e^{i\phi}s$ state there is the time-reversed
$d+e^{-i\phi}s$ state of the same free energy density so that
time-reversal domains may form in a sample. 
Since averaging over time-reversal domains (in the case where
the size of these domains is much smaller than the size of the Josephson 
junction) makes the effects of broken time-reversal symmetry
unobservable, the case of a single time-reversal domain is considered 
here.

Much of this paper studies Josephson tunneling from the $d+e^{i\phi}s$
state to a conventional $s$-wave superconductor, and compares it
with tunneling between a $d+s$ state and an $s$-wave superconductor.
The principal effect of broken time-reversal symmetry is that the 
Josephson junction acquires a characteristic phase which is neither
zero nor $\pi$. The differences in the characteristic phases of two
junctions can be measured in a SQUID experiment.

The symmetry characteristics of the Josephson currents in a 
twinned crystal can be understood in terms of the macroscopic
invariance group of the superconducting state of the twinned crystal.
For the case where the twins are in their time-reversal invariant
$d+s$ state (studied in Ref.~\onlinecite{wal95b}) the macroscopic
invariance group is that of a tetragonal $d_{x^2-y^2}$ superconductor
for odd-symmetry twin boundaries, and it is that of a tetragonal
$s$-wave superconductor for even-symmetry twin boundaries.
Since macroscopic $d$-wave behavior is observed in basal-plane
Josephson tunneling experiments on highly twinned 
crystals~\cite{wo93,ma95b,wo95,va95b,ig94,%
ts94,ts95c,ts95d,ki95b}, 
the twin boundaries for the 
$d+s$ state have to be of the odd-symmetry type~\cite{wal95b}, 
which implies that the $d$-wave component of the order parameter
is approximately continuous across twin boundaries.

In the case of time-reversal symmetry-breaking $d+e^{i\phi}s$ states, 
two types of twin boundaries are consistent with a continuous 
$d$-wave component, namely the odd-symmetry and the
${\cal{T}}\sigma$ twin boundaries.
For odd-symmetry twin boundaries the macroscopic invariance group
is that of a tetragonal $d_{x^2-y^2}$ superconductor, but without the 
time-reversal operation. Detailed calculations confirm, that the 
Josephson tunneling for the $d+s$ state and the $d+e^{i\phi}s$ states 
are indistinguishable, and that they have the predicted macroscopic 
$d$-wave characteristics when the twins are present with equal weights.
For ${\cal{T}}\sigma$ twin boundaries, on the other hand,
the macroscopic invariance group is that of a tetragonal
$d+is$ superconductor. As expected from the macroscopic symmetry, 
in this case  
the $c$-axis Josephson currents cancel only 
partially when averaged over twins, and the 
Josephson junction characteristic phase difference measured in 
a corner SQUID experiment can differ from either zero or $\pi$.

In order to determine the symmetry of the order parameter describing
superconductivity in a single twin (or equivalently, the invariance
group of this superconducting state), it is essential to do
experiments on untwinned single crystals. Experiments on 
twinned crystals determine the macroscopic invariance group of the
twinned crystal, which in general will be different from that
of the superconducting state of an untwinned crystal, and which is
determined by the symmetry of the twin boundaries.

A final section of the paper, stimulated by recent tricrystal-ring 
magnetometry experiments~\cite{ts96} on tetragonal 
{Tl$_2$Ba$_2$CuO$_{6+\delta}$}, proposes and analyzes a ring geometry
which allows $s$ and $d_{x^2-y^2}$ superconductivity to be distinguished
by symmetry arguments only, and which makes unnecessary any assumption
that an $s$-wave superconductor is unlikely to produce a characteristic
phase of $\pi$ in a Josephson junction. The characteristics of 
$d+is$ superconductivity for this geometry are also identified.

\acknowledgements

This work was supported by the Natural Sciences and Engineering Council of
Canada. Useful discussions with R.~C. Dynes, A.~E. Jacobs, and 
C.~C. Tsuei are also acknowledged. 

\appendix

\section{Order Parameter Symmetry in {YBCO}}\label{appa}

On the basis of experimental evidence, this section justifies the assumption 
that the superconducting order parameter in YBCO is a mixture of order
parameters having $d_{x^2-y^2}$ and $s$ wave symmetry. 
It is known~\cite{si95b,si92,ge86,ge87,go87,an90} that the dc Josephson effect
can be used to determine the symmetry of the superconducting order 
parameter. Since YBCO is orthorhombic, the logical first question is:
Which of the irreducible representations of the orthorhombic point
group does the order parameter belong to? 
The argument is straightforward and elementary but we have been
unable to find it in detail elsewhere and therefore present it here.

Consider a Josephson junction between YBCO and lead, the two 
superconductors being characterized by order parameters
$\psi_{\Gamma}$ and $\psi_{\text{Pb}}$, respectively.
Here, $\psi_{\Gamma}$ transforms according to the irreducible
representation $\Gamma$ of the orthorhombic point group $D_{2h}$ of
YBCO and $\psi_{\text{Pb}}$ is assumed invariant under all 
transformations. 
Symmetry considerations show that the lowest order
(in $\psi_{\Gamma}$ and $\psi_{\text{Pb}}$) contribution 
to the Josephson current across a junction will have the form
\begin{equation}\label{a1}
J_{\bf n}^{\Gamma}=iC_{\bf n}^{\Gamma}\left[ 
\psi_{\Gamma}\psi_{\text{Pb}}^*-\psi_{\Gamma}^*\psi_{\text{Pb}}\right] \, ,
\end{equation}
where ${\bf n}$ is the normal to the plane of the junction, and where
$J_{\bf n}$ is the current in direction ${\bf n}$.

Table~\ref{tab1} shows which of the coefficients $C_{\bf n}^{\Gamma}$ are
zero (by symmetry) and which can be nonzero, for each irreducible
representation and for currents along each of the principal 
orthorhombic axes. Since only the $ux^2+vy^2$ 
representation in Table~\ref{tab1}
can have nonzero Josephson currents along all three principal axes, 
and since nonzero Josephson currents have apparently 
been observed experimentally
along all three principal axes in untwinned single crystal 
experiments~\cite{wo93,br94,dy94}, the $ux^2+vy^2$ representation, 
also commonly called the $A_{1g}$ representation,  would appear
to be the only acceptable one.

When the bilinear contribution to the Josephson current is zero, 
however, the biquadratic contribution
\begin{equation}\label{a2}
J_{\bf n}^{(4) \Gamma}=iC_{\bf n}^{(4) \Gamma}\left[ 
\psi_{\Gamma}^2{\psi_{\text{Pb}}^{2*}}
-{\psi_{\Gamma}^2}^*\psi_{\text{Pb}}^2\right] 
\end{equation}
may be nonzero. Here, $C_{\bf n}^{(4) \Gamma}$ is never required to
be zero by symmetry. Close to the critical temperature 
$T_{\text{c,Pb}}\simeq 7.2$K for lead, the biquadratic  
critical current should have a temperature dependence
$J^{(4) \Gamma}\propto|\psi_{\text{Pb}}|^2\propto(T_{\text{c,Pb}}-T)$, 
i.e.\ linear in $T_{\text{c,Pb}}-T$, whereas the bilinear
critical current should have the square root temperature
dependence 
$J^{\Gamma}\propto|\psi_{\text{Pb}}|\propto(T_{\text{c,Pb}}-T)^{1/2}$.
Measurements of the temperature dependence of the $c$-axis critical
current by Dynes and coworkers~\cite{su94,ka95,dy94} clearly
show the square root temperature dependence of the bilinear
current.

We are not aware of measurements which establish experimentally
that the Josephson currents along the $a$ and $b$ axes are of the bilinear, 
as opposed to the biquadratic, type.
An experiment which does this would be of interest.

Note that the biquadratic Josephson current is proportional
to $\sin[2(\phi_{\Gamma}-\phi_{\text{Pb}})]$, 
where $\phi_{\Gamma}-\phi_{\text{Pb}}$ is the phase difference 
between the two superconductors, 
whereas
the bilinear Josephson current is proportional to 
$\sin(\phi_{\Gamma}-\phi_{\text{Pb}})$, 
This difference
means that a SQUID made from Josephson junctions for which
the current is biquadratic would have a maximum
current periodic in the flux with period $\Phi_0/2$ rather
than $\Phi_0$ as is usually the case ($\Phi_0$ is the 
elementary flux quantum). Also, the base frequency for the
ac Josephson effect would be $\omega_{\text{J}}=4eV/\hbar$
rather than $2eV/\hbar$.
For a microscopic calculation of the 
effect for tetragonal systems see  Refs.~\onlinecite{ta94,zh95}. 

If the bilinear character of the Josephson currents 
along the $a$ and $b$ axes of YBCO can be confirmed experimentally, 
then the $ux^2+vy^2$ character of the superconducting order parameter
in YBCO will be established. Of all the different irreducible
representations of the tetragonal point group, only the 
$d_{x^2-y^2}$ and $s$ representations are invariant under the
operations of the orthorhombic point group.
(A different way of seeing this is to note that 
$ux^2+vy^2=d(x^2-y^2)+s(x^2+y^2)$, where the coefficients
$d$ and $s$ can be found in terms of $a$ and $b$.)
Thus the $ux^2+vy^2$ order parameter for YBCO can be considered
to be a combination of $d_{x^2-y^2}$ and $s$ (e.g.\ $x^2+y^2$)
basis vectors
(and no other type of tetragonal symmetry).
These arguments assume a second order transition to the 
superconducting state.

\section{Angular Dependence of Josephson Currents}\label{appb}

Figure~\ref{fig6} shows a Josephson junction between two identical
tetragonal superconductors. The free energy per unit area of the
Josephson junction in zero magnetic field has the form
\begin{equation}\label{appb1}
F=C(\theta,\theta')\cos(\phi-\phi') , 
\end{equation}
where the superconducting order parameters on the two sides of the
junction are $\psi=|\psi|e^{i\phi}$ and $\psi'=|\psi'|e^{i\phi'}$, 
and $\theta$ and $\theta'$ define the orientations of the crystallographic
axes, as shown in the figure.

First assume that the superconductivity is either $s$-wave or 
$d_{x^2-y^2}$-wave. (Our definition of $s$-wave is that the order parameter
is invariant with respect to all operations of the tetragonal point group;
this is sometimes called generalized $s$-wave superconductivity. 
Similarly, our definition of $d_{x^2-y^2}$-wave superconductivity
encompasses generalized $d_{x^2-y^2}$-wave superconductivity.)
Then symmetry considerations show that the coefficients must satisfy the
following conditions:
\begin{equation}\label{appb2}
C(\theta,\theta')=C(\theta',\theta)=C(-\theta,-\theta')=C(\theta+\pi,\theta') .
\end{equation}
Furthermore, for $s$-wave superconductivity
\begin{equation}\label{appb3}
C_s(\theta+\pi/2,\theta')=C_s(\theta,\theta') , 
\end{equation}
whereas for $d_{x^2-y^2}$ superconductivity
\begin{equation}\label{appb4}
C_d(\theta+\pi/2,\theta')=-C_d(\theta,\theta') .
\end{equation}
The use of these constraints yields the following Fourier series 
representations for the coefficients:
\begin{eqnarray}\label{appb5}
C_s(\theta,\theta')&=&\sum_{n,n'}\left[C_{4n,4n'}\cos(4n\theta)\cos(4n'\theta')
\right. \nonumber \\
& &\left. + S_{4n,4n'}\sin(4n\theta)\sin(4n'\theta')\right]  
\end{eqnarray}
for $s$-wave superconductivity, and 
\begin{eqnarray}\label{appb6}
C_d(\theta,\theta')&=&\sum_{n,n'}\left\{C_{4n+2,4n'+2}
\cos[(4n+2)\theta]\cos[(4n'+2)\theta']
\right. \nonumber \\
& &\left. + S_{4n+2,4n'+2}
\sin[(4n+2)\theta]\sin[(4n'+2)\theta']\right\}  
\end{eqnarray}
for $d_{x^2-y^2}$-wave superconductivity. The sums over $n$ and $n'$ are
over all positive integers and zero; also, the coefficients
have the property $C_{i,j}=C_{j,i}$ and $S_{i,j}=S_{j,i}$, where
$i,j$ are $4n$, $4n'$ or $4n+2$, $4n'+2$. 

The first few terms of these expressions are
\begin{equation}\label{appb7}
C_s(\theta,\theta')=C_{0,0}+C_{4,0}[\cos(4\theta)+\cos(4\theta')] 
+\ldots , 
\end{equation}
and 
\begin{equation}\label{appb8}
C_d(\theta,\theta')=C_{2,2}\cos(2\theta)\cos(2\theta')+ 
S_{2,2}\sin(2\theta)\sin(2\theta') +\ldots .
\end{equation}
Note that the expression proposed by Sigrist and Rice~\cite{si92}
for the $d_{x^2-y^2}$ case, namely 
$C_{\text{SR}}(\theta,\theta')=C_{2,2}\cos(2\theta)\cos(2\theta')$, 
has the property $C_{\text{SR}}(\theta,\pi /4)=0$, which is not
required by symmetry and therefore cannot be expected to be true in
general.
The simplified expression given in Eq.~(\ref{appb8}), 
which contains two constants 
$C_{2,2}$ and $S_{2,2}$, does not have this difficulty.
Also, if $S_{2,2}=-C_{2,2}$, the sines and cosines in our
two-term expression Eq.~(\ref{appb8}) can be summed to give
$C_d=C_{2,2}\cos[2(\theta+\theta')]$, which is the expression 
proposed by Tsuei {\it et al.}~\cite{ts95c,ts96} to model
what they call the maximum disorder limit of a grain boundary
Josephson junction.

\begin{table}
\caption{Josephson Currents for Orthorhombic Symmetry.
The value of the critical Josephson current is given for each
irreducible representation $\Gamma$ and for each of the
orthorhombic principal axes, $a$, $b$, and $c$.}
\label{tab1}
\begin{tabular}{c} $
\begin{array}{c|c|c|c}
 \Gamma  &  a  &  b  &  c  \\ \hline
 ux^2+vy^2 & C^1_a & C^1_b & C^1_c \rule{0ex}{3ex} \\
\hspace{2.5em} yz \hspace{2.5em} & \hspace{2.5em} 0 \hspace{2.5em} & 
\hspace{2.5em} 0 \hspace{2.5em} & \hspace{2.5em} 0  \hspace{2.5em}\\
 zx &0&0&0 \\
 xy &0&0&0 \\
 xyz &0&0&0 \\
 x & C^6_a &0&0 \\
 y &0& C^7_b &0 \\
 z &0&0& C^8_c  
\end{array} $
\end{tabular}
\end{table}

\begin{figure}
\caption{Unit cells in neighboring
twins under the assumption that the twin boundary (dashed line)
is a plane of reflection symmetry of the crystal. }
\label{fig1}
\end{figure}

\begin{figure}
\caption{Schematic representation of $d_{x^2-y^2}$- and $s$-wave 
order-parameter components
near odd and even reflection-symmetry twin-boundaries.}
\label{fig5}
\end{figure}

\begin{figure}
\caption{(a) The structure of the interface between a region containing
$(110)$ twin boundaries and a region containing $(\bar{1}10)$ twin
boundaries, as described in Ref.~\protect{\onlinecite{zh93}}.
The rectangles indicate the relative orientations of the orthorhombic
$a$ and $b$ axes in the different twins, and the plus and minus
signs indicate the relative sign of the superconducting order parameter
in the different twins.
The $c$ axis is normal to the 
plane of the figure. Part (b) of the figure is obtained from part (a)
by a rotation of $\pi/2$ about the $c$-axis. }
\label{fig2}
\end{figure}

\begin{figure}
\caption{Corner SQUID geometry for basal-plane tunneling between 
YBCO and lead. $A$ and $B$ indicate basal-plane 
Josephson junctions on two
perpendicular faces of the YBCO crystal, while $I_A$ and $I_B$
denote the corresponding currents. }
\label{fig3}
\end{figure}

\begin{figure}
\caption{
Proposed geometry for ring experiment to test for $s$-wave versus
$d_{x^2-y2}$-wave superconductivity in tetragonal systems.
$S1$ and $S2$ are tetragonal superconductors with their $c$-axis 
normal to the plane of the figure and their basal-plane unit cells 
rotated by an angle $\pi/4$ with respect to each other.
$A$ and $B$ are Josephson junctions; the angle $\alpha$ is chosen
so that $\alpha\neq 0$, $\alpha\neq\pi/4$.} 
\label{fig4}
\end{figure}

\begin{figure}
\caption{Josephson junction between two identical tetragonal
superconductors.}
\label{fig6}
\end{figure}

\end{document}